\documentclass{amsart}
\usepackage{amstext, amssymb, amsthm, amsfonts, latexsym,bbm,tikz}
\usepackage{mathtools}
\usetikzlibrary{matrix}
\usepackage{varioref}
\usepackage{float}
\usepackage{longtable}
\usepackage{todonotes}
\usepackage{bbm}
\usepackage[utf8]{inputenc}
\usepackage[english]{babel}
\usepackage{booktabs}
 
\usepackage{comment}
\usepackage{tikz-cd}
\usepackage{hyperref}
\usepackage[version=3]{mhchem}
\usepackage[most]{tcolorbox}

\theoremstyle{definition}

\theoremstyle{remark}

\theoremstyle{Conjecture/open problem}

\theoremstyle{assumption}

\theoremstyle{conjecture}

\def\1{\mathbf{1}}

\begin{document}

\title[Decision curves, PPV and calibration]{
Why decision curves go above or below treat-all and treat-none: a PPV- and calibration-based guide for clinical prediction models}

\author{Linard Hoessly}
\address{Data Center of the Swiss Transplant Cohort Study, University Basel \& University Hospital Basel, Basel, 4031 Switzerland}

\begin{abstract}
Net benefit is widely used to assess whether clinical prediction models can improve decision-making, but its interpretation is often difficult in practice. In this didactic note, we show how decision curves can be understood through two familiar quantities: positive predictive value and observed event rates in threshold-defined patient groups. We show that a model has positive net benefit compared with treating no one when, among patients classified as positive, the observed event rate exceeds the decision threshold. Similarly, a model outperforms treating everyone when, among patients classified as negative, the observed event rate is below the decision threshold. These results connect decision-curve performance directly to clinically interpretable forms of threshold-specific calibration. We also describe how positive predictive value curves can complement decision curves by showing, at each threshold, the observed event rate among patients selected for action. Together, these perspectives help explain why a prediction model performs better or worse than default strategies and may make decision curve analysis more accessible to clinical audiences. \end{abstract}


\maketitle

\section{Introduction}

Decision-analytic measures play a central role in evaluating the impact of clinical prediction models \cite{steyerberg2019clinical,Steyerberg2010_scaled_2}. In the following, we focus on net benefit (NB), which has no need to estimate cost or utility, and has become widely used \cite{vancalster2024performanceevaluationpredictiveai}. In particular, while ealier reporting guidelines like TRIPOD \cite{Collins2015} mention NB implicitly, more recent ones like TRIPOD-Cluster \cite{Debray2023}, TRIPOD+AI \cite{Collins2024}, or TRIPOD-SRMA \cite{Snell2023} explicitely recommend or at least mention decision curve analysis respectively NB. In practice, NB is typically compared with the default strategies treat-all and treat-none for both prognostic and diagnostic models.

Despite extensive methodological work, tutorials, and applied use \cite{Vickers2006,Vickers2008,cal_net_benefit,vancalster2024performanceevaluationpredictiveai,Talluri2016,Rousson2011,VanCalster2018_misuse_netben,Kerr2016}, interpreting NB can still be difficult \cite{VanCalster2018_misuse_netben}.  In particular, it is often unclear what it means, clinically and practically, when a model falls below treat-none or treat-all at a given threshold. In this paper, we address this question by re-expressing these comparisons in terms of positive predictive value (PPV) and threshold-specific observed risk.

Note that NB is proper as a scoring rule \cite{Pepe2014}: in expectation, it is maximized by the true risks (i.e., perfect predictions) and cannot be improved upon \cite{scoring_rule}. Like the Brier score what we observe in practice is not the expectation but observed values. Hence observed NB is a realisation of a random variable and is therefore subject to sampling variability \cite{Hoessly2026,Sorahan1994}. For this purpose, it is advisable to display uncertainty, for example via bootstrap confidence bands or, at minimum, threshold-specific uncertainty summaries \cite{steyerberg2019clinical,Hoessly_cje_key2026}.
NB is analytically linked to PPV \cite{zhou_relationship_2021}. Simulation studies have shown that systematic miscalibration can substantially reduce net benefit \cite{cal_net_benefit}, which is consistent with the properness of the underlying scoring-rule formulation \cite{scoring_rule}.

Motivated by the practical difficulty of interpreting net benefit, we revisit a known algebraic link between net benefit and positive predictive value and use it to provide a clinically intuitive explanation of decision-curve results. We also develop a threshold-specific calibration interpretation of comparisons with the default strategies of treat-none and treat-all. Specifically, we show that a model outperforms treat-none at a given threshold when the observed event rate among patients classified as positive exceeds that threshold, and that it outperforms treat-all when the observed event rate among patients classified as negative is below that threshold. These equivalences translate decision-analytic statements into quantities that are often easier for clinicians to understand. Finally, we propose PPV curves as a practical visual companion to decision curves. We do not introduce a new decision metric; rather, we offer a didactic framework and diagnostic visualisation to make net benefit easier to interpret, explain, and troubleshoot in applied clinical prediction research.

\subsection{The prediction setting and classification evaluation}
Consider $n$ patients. For patient $i$, a model predicts an event risk $p_i\in[0,1]$, and we observe the outcome $y_i\in\{0,1\}$. The true (unknown) risk is $q_i\in[0,1]$, with perfect prediction given by $p_i=q_i$ \cite{harrell2015regression,Rufibach2010}.

\medskip
\noindent
Given a decision threshold $t$, patients with $p_i\ge t$ are classified as positive and those with $p_i<t$ as negative, yielding counts of true positives $TP(t)$, true negatives $TN(t)$, false positives $FP(t)$, and false negatives $FN(t)$. Classification performance is available as
$(TP(t),FP(t),TN(t),FN(t))$. However, most popular classification evaluation measures consist of only one value summarising a specific aspect of performance \cite{hand_classification}. Note that in a dataset the numbers of events and non-events are fixed,
\begin{equation}\label{eq_n_1_n_0}
TP(t)+FN(t)=n_1,\qquad FP(t)+TN(t)=n_0,
\end{equation}
where we denote the event fraction by $\pi=\frac{n_1}{n}$ and the selection rate by $s_t := (TP(t) + FP(t))/n$.
\subsubsection{Net benefit}
The NB at threshold $t$ is then defined as \cite{Vickers2008}
\begin{equation}\label{Net_nen} 
NB(t) \;=\; \frac{TP(t)}{n} - \frac{FP(t)}{n}\cdot\frac{t}{1-t},\quad t\in (0,1).
\end{equation}
This definition can be interpreted as the true positive rate adjusted by a weighted penalty for false positives, where the weight $\tfrac{t}{1-t}$ reflects the relative harm of a false positive compared to the benefit of a true positive at threshold $t$.

In medicine, prediction models are typically judged in a clinically plausible subrange of $(0,1)$ \cite{Vickers2019} against the simple strategies of treat-none and treat-all.  
A model outperforms treat-none when its net benefit is greater than zero, and it outperforms treat-all when its net benefit is higher than that of the treat-all strategy \cite{Vickers2006}. 
As a word of caution, NB is a utility calculation that becomes arbitrary whenever the threshold $t$ is not anchored to real harms/benefits for the specific action and setting. Note that as it is a one-dimensional summary, two models can have equal net benefit despite different $(TP(t),FP(t))$:
\[
NB_1(t)=NB_2(t)
\quad\Longleftrightarrow\quad
TP_1(t)-TP_2(t)=\frac{t}{1-t}\bigl(FP_1(t)-FP_2(t)\bigr).
\]

\subsubsection{PPV and PPV curves}

PPV is the probability that a patient with a positive test truly has the disease, i.e., 
 the proportion of true positives among all positives. In the machine learning literature it is also called precision \cite{hand_classification}.
In practice, PPV indicates how much confidence a clinician can place in a positive test, making it central to the evaluation of diagnostic tools and treatment decisions \cite{Altman102}.
As in the case of NB curves, we consider PPV curves where we have a curve along $(PPV(t),t)$, where
$$PPV(t):=\begin{cases}\tfrac{TP(t)}{TP(t)+FP(t)}\quad &\text{ if }TP(t)+FP(t)\neq 0\\
0&\text{ otherwise},
\end{cases} \quad t\in (0,1)
$$

\subsection{Quick reference for practitioners}
\hfill\break
\begin{table}[h!]
\centering
\renewcommand{\arraystretch}{1.25}
\begin{tabular}{p{4.2cm}p{4.2cm}p{6.0cm}}
\toprule
\textbf{Question} & \textbf{Equivalent condition} & \textbf{Practical meaning} \\
\midrule

Is the model better than treat-none at threshold \(t\)? 
&
\(\bar Y_{\ge t}>t\)
&
Among patients classified positive, the observed event rate must exceed the treatment threshold. Otherwise, the flagged group is not high-risk enough to justify intervention relative to treating nobody. \\

Is the model better than treat-all at threshold \(t\)? 
&
\(\bar Y_{<t}<t\)
&
Among patients classified negative, the observed event rate must be below the treatment threshold. Otherwise, withholding intervention from this group is not justified. \\

How do PPV curves help interpret decision curves? 
&
\(\mathrm{PPV}(t)=\bar Y_{\ge t}\)
&
 PPV at threshold $t$ describes the following: among patients flagged positive, what fraction truly has the outcome.\\

What does poor decision-curve performance mean? 
&
\(\bar Y_{\ge t}\le t\) or \(\bar Y_{<t}\ge t\)
&
Falling below treat-none or treat-all at $t$ has a direct interpretation in terms of threshold miscalibration at $t$. \\

\bottomrule
\end{tabular}
\caption{Practical connections between net benefit, threshold-specific event rates, and PPV. Here \(\bar Y_{\ge t}\) denotes the observed event rate among patients with predicted risk \(p\ge t\), and \(\bar Y_{<t}\) the observed event rate among patients with \(p<t\).}
\label{tab:nb_ppv_connections}
\end{table}

\subsection{Related literature}

Several introductions and tutorials explain how to use and interpret net benefit (NB) and decision curve analysis in applied clinical research \cite{VanCalster2018_misuse_netben,Kerr2016,Mijderwijk2021,Kohn2025,Vickers2016}. Our focus is different: we aim to make comparisons with treat-none and treat-all easier to interpret by expressing them through PPV and threshold-specific observed event rates.

The algebraic link between NB and PPV has been noted previously \cite[p. 12]{zhou_relationship_2021}. Here, we use this link to give explicit conditions under which a model outperforms the default strategies in terms of PPV. The effect of miscalibration on NB has also been studied through simulations \cite{cal_net_benefit}, and it was shown that so-called moderate calibration guarantees non-harmful
decision making in expectation, meaning that NB is not below the default
strategies of treat-none and treat-all \cite{VanCalster2016}. We complement this by showing that moderate calibration is sufficient but not necessary for nonharmful decision making, and commenting on other aspects of moderate calibration. In particular, our results are  
threshold-specific empirical conditions which explain, at each threshold, why
a decision curve lies above or below the default strategies. 

Finally, we propose PPV curves as a visual companion to decision curves. Related displays include classification plots, which show true-positive and true-negative rates across thresholds \cite{Kerr2016,Verbakel2020}, and lift charts, which summarize enrichment among patients ranked by predicted risk \cite{rezavc2011measure}. In contrast, PPV curves show the observed event rate among patients classified as positive at each decision threshold, giving a direct visual explanation of the corresponding decision-curve results.
\subsection*{Acknowledgements}
We thank Matthew Parry, Tinh-Hai Collet, Lucia de Andres, Julien Vionnet, Eveline Daetwyler, Simon Schwab and Louis Faul for helpful discussions and feedback.
\subsection*{AI use}
During the preparation of this manuscript, we used  GPT-4o
 for minor language edits as well as for latex support aiming to enhance readability. After
using it, we reviewed and edited the content as needed
and take full responsibility for its content.

\section{Understanding fundamental conditions for net benefit through PPV and calibration}\label{sec_und_nb_ppv}
\subsection{Interpreting net benefit through PPV curves}\label{subsec:mapping}
The following formula expresses PPV through NB and $s_t$ \cite{zhou_relationship_2021}
\begin{equation}\label{conn_PPV_NB}
PPV(t) =\begin{cases}\ \tfrac{n \, NB(t)}{TP(t) + FP(t)}(1 - t) + t\quad &\text{ if }TP(t)+FP(t)\neq 0\\
0&\text{ otherwise},
\end{cases} \quad t\in (0,1).
\end{equation}

This identity links two perspectives: the decision-analytic view through net benefit and the predictive view through PPV.  
Although there is no one-to-one correspondence between $NB(t)$ and $PPV(t)$, they each place constraints on the other, as outlined for PPV in Appendix \ref{bound_NB}.  
In what follows, we explore how this relationship can be used to derive equivalences between model comparisons to treat all and treat none (detailed mathematical derivations are in Appendix $\S$ \ref{A_ppv}). These can easily be visualised along PPV curves as in $\S$ \ref{vis}, enabling direct visual complements of NB comparisons. 
\begin{itemize}
    \item \textbf{$NB(t)$ better than treat-none means the proportion of true positives among all positives is bigger than t, $t\in(0,1)$:}
\begin{equation}\label{NB_treat_none}
    NB(t) > 0 
    \quad \Longleftrightarrow \quad 
    PPV(t) > t.
\end{equation}
    For instance, at $t = 0.1$, the model must achieve $PPV(t) > 0.1$ for positive NB.
    
    \item \textbf{$NB(t)$ outperforms treat-all at $t$, $t\in(0,1)$:}
\begin{equation}\label{NB_treat_all}
    NB(t) > \pi-(1-\pi)\tfrac{t}{1-t} 
    \quad \Longleftrightarrow PPV(t) > \frac{\pi - t}{s(t)} + t,
\end{equation}
    where the right-hand side depends on the number of patients classified positive.  
   
  \end{itemize}

\subsection{Interpreting net benefit through decision-threshold calibration}\label{subsec:calibration}
Calibration measures how far observed event rates are from predicted event rates. In this section, we derive equivalences between NB comparisons to treat all and treat none.

Let $\bar{Y}_{\ge t}$ denote the observed event rate among individuals whose predicted probability exceeds $t$, $\bar{Y}_{< t}$ the observed event rate among those whose predicted probability is below $t$ and let $\bar{p}_{\ge t}$,$\bar{p}_{< t}$ denote the average predicted probabilities over the corresponding groups. In particular, if the predictions are well-calibrated, we can assume that we observe
$$
\bar{Y}_{\ge t}\approx \bar{p}_{\ge t},\quad \bar{Y}_{< t}\approx \bar{p}_{< t},
$$
and by definition (except in degenerate cases), we have
$\bar{p}_{\ge t} > t,\bar{p}_{< t} < t$. The above checks are small extensions of calibration in the large checks like $\bar{Y}\approx \bar{p} $  \cite{steyerberg2019clinical}. Hence
two simple decision-threshold calibration checks are
\begin{enumerate}
\item $
\bar{Y}_{\ge t} > t$
\item $\bar{Y}_{< t} < t.
$
\end{enumerate}

Surprisingly, these are equivalent to the following, where we assume that there are some cases classified positive and negative under threshold $t\in (0,1)$, i.e. $1>s_t>0$ (detailed derivations are given in Appendix \ref{A2})
\begin{itemize}
\item \textbf{$NB(t)$ better than treat-none means $
\bar{Y}_{\ge t} > t$, i.e., observed event rate among individuals whose predicted probability exceeds $t$ is in the right direction:}
We can write \begin{equation}\label{eq_cal_decomp1}
NB(t) = \frac{s_t}{1 - t}\,(\bar{Y}_{\ge t} - t).
\end{equation}
Thus $NB(t) > 0$ exactly when $\bar{Y}_{\ge t} > t$.  
More generally, net benefit can be viewed as the product of two factors:
a calibration surplus $(\bar{Y}_{\ge t} - t)$ and a selection-rate multiplier $s_t / (1 - t)$, which scales this surplus according to how many individuals are classified as positive.
 \item \textbf{$NB(t)$ outperforms treat-all at $t$ means $\bar{Y}_{< t} < t$, i.e., observed event rate among individuals whose predicted probability is below $t$ is in the right direction:}

\begin{equation}\label{eq_cal_decomp2}
NB(t) > NB_{\mathrm{all}}(t)
\quad\Longleftrightarrow\quad
\bar{Y}_{<t} < t.
\end{equation}
A prediction model beats treat all at threshold $t$ exactly when the patients not treated
(those with predicted risk below $t$) truly have an observed event rate below $t$.
In other words, it is beneficial to withhold treatment from the below-threshold group only if that group is
indeed low risk in the observed data.
\end{itemize}

\subsection{Example visualisation of PPV curves}\label{vis}
To illustrate PPV curves, we present examples of prediction models together with their decision curve analysis (net benefit) and the corresponding PPV curves, complemented with plots of the distribution of predictions. Calibration plots as in Section~\ref{subsec:calibration} are provided in Appendix~\ref{vis_cal}, as calibration plots are confirmatory here since this follows by the analytic threshold-specific interpretation.

Decision curves were produced with the \texttt{dcurves} package \cite{dcurves}. We implemented an analogous routine to compute and plot PPV curves. Each decision curve figure shows the model-specific NB curves as well as the treat-all and treat-none reference curves. Each PPV figure includes at least 3 curves and lines, whose color is shared across all figures for consistency:
\begin{itemize}
\item the treat-none reference (the main diagonal), consistent with \eqref{NB_treat_none};
\item the PPV curve;
\item for each model, a dotted ``treat-all comparison'' curve (same color as the model's PPV curve), obtained from the right-hand side of \eqref{NB_treat_all}.
\end{itemize}
Colors are matched across figures: a model's NB curve, PPV curve, and dotted treat-all comparison curve share the same color.

\begin{tcolorbox}
\textbf{Interpretation.} For a given threshold $t$:
\begin{itemize}
\item if the model's PPV curve lies above the main diagonal, then $NB(t)>0$ (better than treat-none);
\item if the model's PPV curve lies above the dotted treat-all comparison curve, then $NB(t)>NB_{\mathrm{all}}(t)$ (better than treat-all).
\end{itemize}
\end{tcolorbox}

\begin{figure}[ht]
    \centering
    \includegraphics[width=1.3\textwidth]{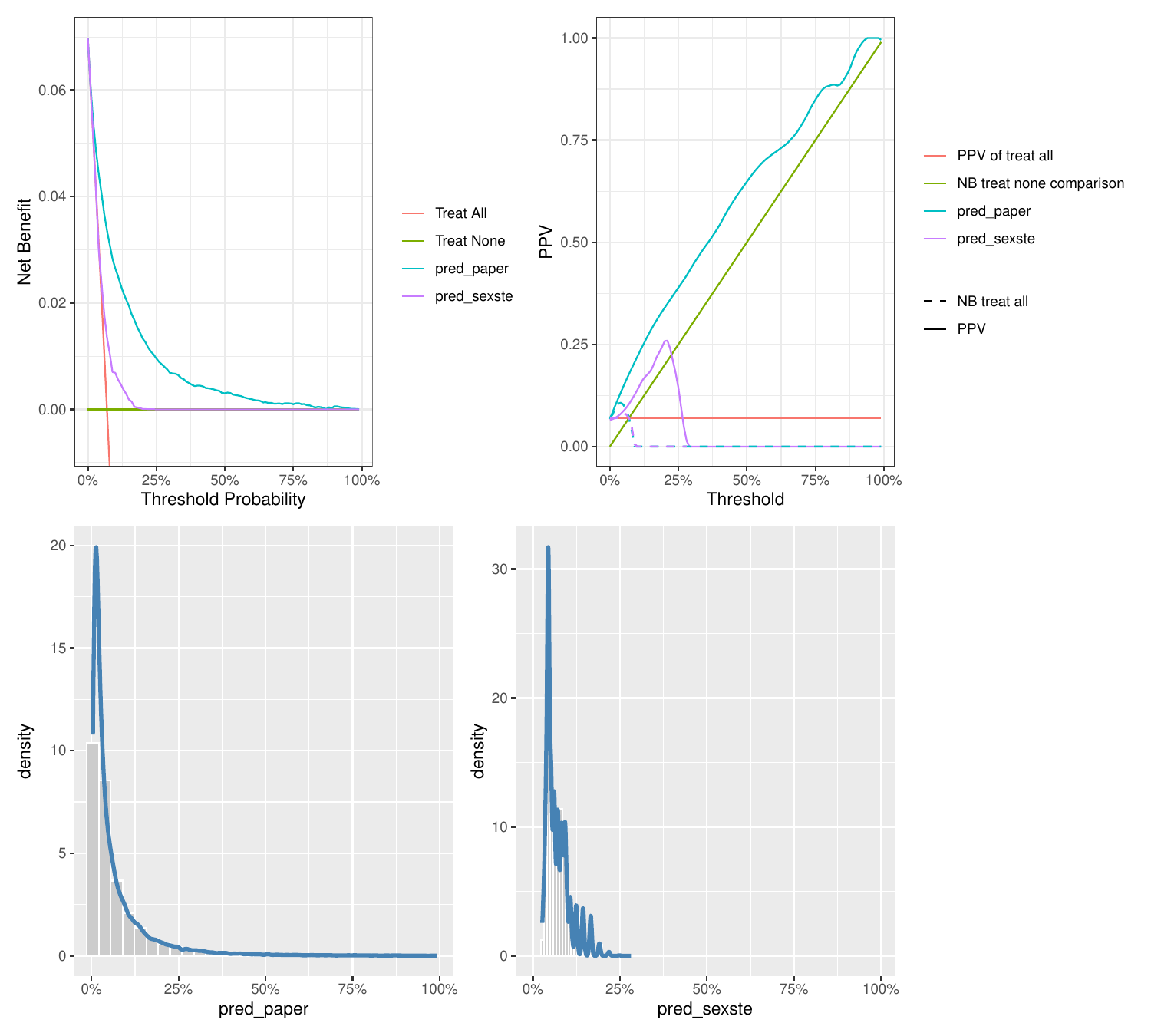}
    \caption{Net benefit curve with corresponding PPV curve for GUSTO-I.}
    \label{fig_1}
\end{figure}

\begin{figure}[ht]
    \centering
    \includegraphics[width=1.3\textwidth]{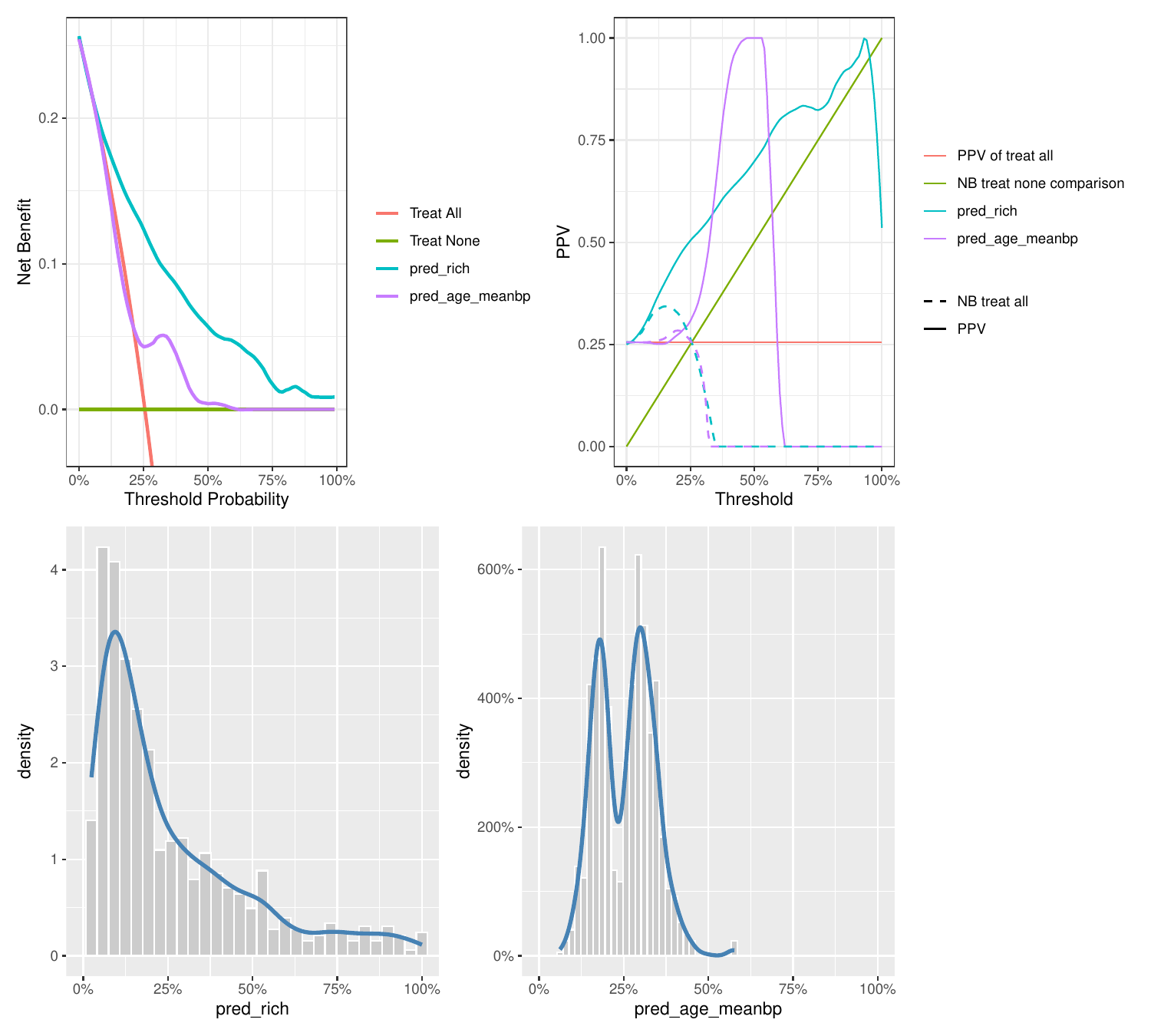}
    \caption{Net benefit curve with corresponding PPV curve for SUPPORT.}
    \label{fig_2}
\end{figure}

We use the following illustrative examples:
\begin{itemize}
\item Figure~\ref{fig_1} (GUSTO-I \cite{Lee1995}, complete cases): we compare (i) a ``published-style'' logistic regression including flexible effects for age ($\mathrm{ns}(age,\mathrm{df}=4)$) and heart rate ($\mathrm{ns}(pulse,\mathrm{df}=3)$), Killip class, winsorised systolic blood pressure ($\text{sysbp}_w=\min(\text{sysbp},120)$), infarct location (\texttt{miloc}), and an age$\times$Killip interaction; with (ii) a parsimonious logistic regression using sex and ST-elevation burden (\texttt{ste}).

\item Figure~\ref{fig_2} (SUPPORT \cite{Knaus1995}, complete cases): we compare (i) a richer logistic regression model including flexible effects for age, mean arterial pressure, and creatinine (restricted cubic splines with 5 knots: $\mathrm{rcs}(\cdot,5)$), together with sex, disease class (\texttt{dzclass}), number of comorbidities (\texttt{num.co}), coma score (\texttt{scoma}), white blood cell count (\texttt{wblc}), and respiratory status (\texttt{resp}); with (ii) a simple logistic regression using only age and mean arterial pressure.
\end{itemize}
\color{black}
The advantages of PPV curves are illustrated in Figures~\ref{fig_1}--\ref{fig_2}. At higher thresholds, net benefit curves often shrink toward zero, making comparisons with treat-all and treat-none harder to read. PPV curves make the same comparisons immediate: the diagonal $PPV(t)=t$ corresponds to the treat-none boundary, and the dotted curve to the treat-all boundary. Wherever a models PPV curve lies above these references, its net benefit is above the corresponding default strategy at that threshold. Moreover, PPV provides clinically relevant information that is not explicit in NB: it is the probability that a patient flagged positive truly has the outcome. In both GUSTO-I and SUPPORT, the richer models remain above the reference curves over a wider range of thresholds than the simpler models, which lose separation more quickly.

R version 4.4.3 was used for the illustration, where we loaded the GUSTO-I \cite{Lee1995} and SUPPORT \cite{Knaus1995} dataset  through Hmisc \cite{Hmisc}.

\color{black}

\subsection{Analytical arguments on calibration and net benefit}
We review the observations that miscalibration can substantially reduce clinical utility and may even lead to
clinical harm, i.e.\ net benefit below the treat-all or treat-none strategies \cite{cal_net_benefit}.Two
 failure modes were highlighted: systematic overestimation can yield $NB(t)<0$ for thresholds $t>I$ (worse than treat-none),
whereas systematic underestimation can yield $NB(t)<NB_{\text{all}}(t)$ for thresholds $t<I$ (worse than treat-all).
Both effects can be explained by the observations in Section~\ref{subsec:calibration}. For convenience we briefly
go through the arguments below.

\medskip
\noindent\textbf{Overestimation.} If risks are systematically overestimated, some individuals with true risk below $t$
are pushed above the threshold and treated. This dilutes the treated set, so its observed event rate can fall below the
threshold $\bar Y_{\ge t}<t$, implying by \eqref{eq_cal_decomp1}
that $NB(t)<0$ (NB worse than treat-none). This is more likely for
high thresholds, e.g.\ $t>I$ \cite{cal_net_benefit}.

\medskip
\noindent\textbf{Underestimation.} If risks are systematically underestimated, some individuals with true risk above $t$
are pushed below the threshold and left untreated. This inflates the risk of the untreated set, potentially yielding
$\bar Y_{<t}\ge t$, which is equivalent to
$
NB(t)<NB_{\mathrm{all}}(t)$ by \eqref{eq_cal_decomp2}. This is most likely for low thresholds,
e.g.\ $t<I$, matching the simulations in \cite{cal_net_benefit}.

\medskip
In summary, these failures correspond to violations of the calibration inequalities
$\bar Y_{\ge t}>t$ (treated set) and $\bar Y_{<t}<t$ (untreated set), with their impact scaled by $s_t$ and $1-s_t$,
respectively.

\section{Advanced considerations}
\subsection{A further decomposition of net benefit}

Define the selected-set calibration error
\[
\Delta_t := \bar Y_{\ge t}-\bar p_{\ge t}
= PPV(t)-\bar p_{\ge t},
\]
where $\bar p_{\ge t}$ is the mean predicted risk and $\bar Y_{\ge t}$ the observed event rate among those with $\hat p\ge t$.
Then
\[
NB(t)=\frac{s_t}{1-t}\,(\bar Y_{\ge t}-t)
=\frac{s_t}{1-t}\Bigl((\bar p_{\ge t}-t)+\Delta_t\Bigr).
\]
This decomposes net benefit into 
\begin{itemize}
\item an enrichment term $\frac{s_t}{1-t}(\bar p_{\ge t}-t)$, which adds how far the average
predicted risk in the selected set exceeds $t$, and
\item a calibration term $\frac{s_t}{1-t}\Delta_t$, which adds a bonus
if $\Delta_t>0$ and a penalty if $\Delta_t<0$. 
\end{itemize} 
Hence for a very large number of observations with perfect predictions, leading to a calibration term of roughly zero and positive enrichment means that perfect predictions for net benefit can be expected to be positive by large sample argument (i.e., law of large numbers or central limit theorem \cite{georgii2008stochastics}).

 Also note that the factor $s_t/(1-t)$ implies that miscalibration matters most when many
patients are selected (large $s_t$) and at lower thresholds.
\subsection{Understanding model comparisons in NB through PPV and calibration}
In the following, we derive conditions for a model to be superior in net benefit at threshold $t$ and express them equivalently in terms of PPV and calibration. These interpretations are less direct than those in Section~\ref{sec_und_nb_ppv}, but the PPV-based condition is particularly useful in practice because it can be visualized easily via PPV curves.

\subsubsection{Comparing two models in NB through PPV}  
    Let models $M_1$ and $M_2$ be given with their net benefits denoted
    \[
    NB^{(j)}(t)= \frac{TP^{(j)}(t)}{n}- \frac{t}{1-t}\frac{FP^{(j)}(t)}{n}, \quad j=1,2.
    \]
    Then $M_1$ outperforms $M_2$ in terms of $NB(t)$ if and only if we have
\begin{equation}\label{eq_PPV_mNB}
    NB^{(1)}(t) > NB^{(2)}(t) 
    \quad \Longleftrightarrow \quad
    PPV^{(1)}(t) > t + (1 - t)\frac{n \cdot NB^{(2)}(t)}{TP^{(1)}(t) + FP^{(1)}(t)}.
\end{equation}
    Graphically, this means that at each threshold $t$, $M_1$'s PPV curve must lie above the reference curve defined by $M_2$'s net benefit, the right hand side of \eqref{eq_PPV_mNB}.

\subsubsection{Higher net benefit as a calibration condition}
For each model $m\in\{1,2\}$, assume that at threshold $t\in(0,1)$ there are both positive and negative classes. Then denote
\[
s_{m,t}:=\frac{TP_m(t)+FP_m(t)}{n},\qquad
\bar Y_{m,\ge t}:=\frac{TP_m(t)}{TP_m(t)+FP_m(t)},\qquad
\bar Y_{m,<t}:=\frac{FN_m(t)}{TN_m(t)+FN_m(t)},
\]
from which we can derive the following (see Appendix \ref{A3}),
\[
NB_m(t)=\frac{s_{m,t}}{1-t}\,(\bar Y_{m,\ge t}-t)
\;=\;
NB_{\mathrm{all}}(t)+\frac{1-s_{m,t}}{1-t}\,(t-\bar Y_{m,<t}).
\]
Hence,
\begin{equation}\label{eq_cal_mNB}
\begin{aligned}
NB_1(t) &> NB_2(t)\\
&\Longleftrightarrow\ 
  s_{1,t}(\bar Y_{1,\ge t}-t)
  > s_{2,t}(\bar Y_{2,\ge t}-t)\\
&\Longleftrightarrow\ 
  (1-s_{1,t})(t-\bar Y_{1,<t})
  > (1-s_{2,t})(t-\bar Y_{2,<t}).
\end{aligned}
\end{equation}
That is, higher NB corresponds to a larger above-threshold (or, equivalently, below-threshold) calibration margin, weighted by the fraction treated (or spared).

Furthermore, we note that if $s_{1,t}=s_{2,t},t\in(0,1)$, then the equivalences in \eqref{eq_cal_mNB} simplify to
\[
NB_1(t)>NB_2(t)\quad\Longleftrightarrow\quad \bar Y_{1,\ge t}>\bar Y_{2,\ge t}\quad\Longleftrightarrow\quad  \bar Y_{1,<t}
  < \bar Y_{2,<t},
\]
which are again the observed event rates from $\S$ \ref{subsec:calibration}.

\color{black}

\section{Reviewing the connection of moderate calibration to NB}

\subsection{Why does moderate calibration imply nonharmful decision making hold?}
Denote by $\bar{Y}_{i:p_i=r}$ the mean over all outcomes where the predictions equal $r$.
Then, moderate calibration as in \cite{VanCalster2016} can be expressed as Follows:For any $r\in[0,1]$ that is predicted at least once it holds that
\begin{equation}
\mathbb{E}(\bar{Y}_{i:p_i=r})=r.\label{equation_mod_cal}
\end{equation}
\cite{VanCalster2016} showed that moderate calibration
implies nonharmful decision making in expectation.

For moderate calibration it follows that if there is a prediction above or equal $t$, we have
\[
\mathbb{E}(\bar{Y}_{\ge t})
\ge t,
\]
which, from \eqref{eq_cal_decomp1} implies that we are nonharmful when compared to treat none.
To see this let $r_1,\cdots,r_k$ be the $k$ different values predicted above or equal to $t$ for $1>t>0$, and with each taken $n_1,\cdots,n_k$ times. Then we can rewrite the expectation by \eqref{equation_mod_cal} and the linearity of the expectation as
\[
\mathbb{E}(\bar{Y}_{\ge t})=\mathbb{E}(\tfrac{\sum_{j=1}^kn_j\bar{Y}_{i:p_i=r_j}}{\sum_{j=1}^kn_j})=\tfrac{\sum_{j=1}^kn_j\mathbb{E}(\bar{Y}_{i:p_i=r_j})}{\sum_{j=1}^kn_j}=\tfrac{\sum_{j=1}^kn_jr_j}{\sum_{j=1}^kn_j}\ge \min_{1\leq j\leq k}r_j.
\ge t.
\]
An analogous argument shows that under moderate calibration $\mathbb{E}(\bar{Y}_{<t}) \leq t$, which again by the corresponding statement \eqref{eq_cal_decomp2} implies that we are nonharmful when compared to treat all.

However, note that these implications are entirely theoretical as they concern statements that hold in expectation.
\subsection{What does this mean in practice?}
The equivalence of the calibration inequalities and improvement to treat all and treat none for net benefit in $\S$ \ref{subsec:calibration} are analytic. They hold for observed quantities, and by taking expectations on both sides they can be translated to a theoretical equivalence in expectation. However, we note that
\begin{itemize}
\item  it is typically impossible to conclude in practice that moderate calibration holds as it is a property in expectation. This is clear since whenever we have numerical variables in prediction models, the variable value combinations often lead to unique predicted probabilities \cite{VanCalster2016}.
\item observed nonharmful decision making does not imply theoretical nonharmful decision making and the other direction, which holds by the difference between theoretical and observed clinical utility through net benefit. What we observe in practice is not the expectation but observed values \cite{Hoessly2026,Sorahan1994}.
\item moderate calibration, while nonharmful overall, can be harmful in the subgroups it comprises. An example is provided in Appendix $\S$ \ref{app:moderate-calibration-subgroups}.
\item moderate calibration is sufficient but not necessary for nonharmful decision making. An example is provided in Appendix $\S$ \ref{app:moderate-calibration-not-necessary}.
\end{itemize}

\section{Discussion}

PPV, decision-threshold calibration, and net benefit (NB) describe model performance from different but complementary perspectives. In this paper, we have shown that these perspectives are closely linked through comparisons of NB with the default strategies of treat-none and treat-all.

PPV, the probability that a patient classified as positive at a given threshold truly has the outcome, is clinically intuitive and answers the question: ``If the model recommends action, how likely is it to be correct?'' PPV curves therefore provide a useful companion to decision curves. They preserve the threshold-based clinical interpretation of NB while often making the comparison with default strategies easier to understand. In particular, superiority over treat-none or treat-all can be assessed by whether the PPV curve lies above or below the corresponding threshold-specific reference curve. This may be especially helpful at higher thresholds, where NB curves often approach zero \cite{cal_net_benefit,Kerr2016,Vickers2016}, whereas PPV curves can still show whether meaningful separation from the threshold remains. In this sense, decision curves show which strategy is preferred, and PPV curves help explain why.

Decision-threshold calibration provides a second, complementary interpretation. A model performs worse than treat-none or treat-all at threshold $t$ if and only if the observed event rate in the corresponding subgroup is on the wrong side of $t$, as described in \S\ref{subsec:calibration}. For example, if a model performs worse than treat-none at threshold $t$, then
\[
\bar Y_{\ge t} < t.
\]
That is, among patients with predicted risk at least $t$, the observed event rate is below the threshold required to justify action. Such deviations may arise from sampling variability \cite{Hoessly2026,Sorahan1994}, but they may also indicate systematic miscalibration and thus motivate recalibration or model updating.

Because NB and decision curve analysis are now widely recommended in studies of clinical prediction models \cite{Collins2015,Debray2023,Collins2024,Snell2023}, we hope these connections make NB easier to interpret and apply. For practitioners, the practical implications are straightforward: report NB over clinically relevant thresholds, consider adding PPV curves as an explanatory companion, add confidence bands to estimate uncertainty, and inspect threshold-specific observed event rates above and below the threshold when a model performs poorly relative to treat-none or treat-all.

Finally, we note that our focus has been on the NB of the treated. In some settings, however, the NB of the untreated may be more relevant \cite{Rousson2011}. Parallel results can then be derived using negative predictive value \cite{Akobeng2007} together with decision-threshold calibration, leading to analogous visualisations and interpretations for the untreated group.

\color{black}

\bibliographystyle{plain}

 \bibliography{pred_references} 
 
 \newpage
 \appendix
\section{Mathematical derivations}

\subsection{Bounds on PPV implied by net benefit}\label{bound_NB}

For a fixed incidence $I$, the fractions of true and false positives satisfy
\[
\frac{TP(t)}{n} \in [0, I], 
\qquad 
\frac{FP(t)}{n} \in [0, 1 - I].
\]
From these constraints, one can derive sharp bounds for $PPV(t)$ given $NB(t)$:
\[
PPV(t) \in
\begin{cases}
\Biggl[
\max\!\left\{
\frac{NB(t)}{NB(t)+\tfrac{1}{t}(I-NB(t))}(1-t)+t,\ 
\frac{NB(t)}{NB(t)+\tfrac{1}{1-t}(1-I)}(1-t)+t
\right\},
\ 1\Biggr], 
& \text{if } NB(t) > 0, \\[1.2em]
\{0,\,t\}, 
& \text{if } NB(t) = 0, \\[1.2em]
\Biggl[\,0,\ 
\min\!\left\{
\frac{NB(t)}{-NB(t)+\tfrac{1}{t}I}(1-t)+t,\ 
\frac{NB(t)}{-NB(t)+\tfrac{1}{1-t}(1-I+NB(t))}(1-t)+t
\right\}\Biggr],
& \text{if } NB(t) < 0.
\end{cases}
\]
In words:
\begin{itemize}
    \item If $NB(t) > 0$, $PPV(t)$ must exceed a threshold depending on $I$ and $NB(t)$.
    \item If $NB(t) = 0$, $PPV(t)$ equals either $0$ or $t$.
    \item If $NB(t) < 0$, $PPV(t)$ is bounded above by a function of $I$ and $NB(t)$.
\end{itemize}
Thus, while $NB(t)$ does not uniquely determine $PPV(t)$, it restricts its feasible range.
\section{Mathematical derivations for PPV}\label{A_ppv}
\subsection{Outperforming treat none}\label{A_ppv_none}

Assume $s_t>0$ and $t\in(0,1)$. Then $PPV(t):=\frac{TP(t)}{TP(t)+FP(t)}$, $s_t:=\frac{TP(t)+FP(t)}{n}$, and we can rewrite $TP(t),FP(t)$ as
$$TP(t)=s_t n\,PPV(t), \quad FP(t)=s_t n\,(1-PPV(t)),$$
and hence
\[
NB(t)=\frac{TP(t)}{n}-\frac{t}{1-t}\frac{FP(t)}{n}
=s_t\Bigl(PPV(t)-\frac{t}{1-t}\bigl(1-PPV(t)\bigr)\Bigr)
=\frac{s_t}{1-t}\bigl(PPV(t)-t\bigr).
\]
With that we get
\[
NB(t)>0 \quad\Longleftrightarrow\quad PPV(t)>t.
\]

\subsection{Outperforming treat all}\label{A_ppv_all}
The net benefit of treating everyone is
\[
NB_{\mathrm{all}}(t)=\pi-(1-\pi)\frac{t}{1-t},
\qquad \text{ with } \pi=\frac{n_1}{n}.
\]
Using \eqref{conn_PPV_NB}, we rewrite
\[
NB(t)-NB_{\mathrm{all}}(t)
=\frac{s_t}{1-t}\bigl(PPV(t)-t\bigr)-\Bigl(\pi-(1-\pi)\frac{t}{1-t}\Bigr)>0
\]
by multiplying by $(1-t)>0$ and rearranging, giving
\[
s_t\bigl(PPV(t)-t\bigr)>\pi-t
\quad\Longleftrightarrow\quad
PPV(t)>\frac{\pi-t}{s_t}+t.
\]
\section{Complementing visualisation to equivalence calibration and NB from $\S$ \ref{subsec:calibration} in the settings described in $\S$ \ref{vis}}\label{vis_cal}
\begin{figure}[ht]
    \centering
    \includegraphics[width=1.3\textwidth]{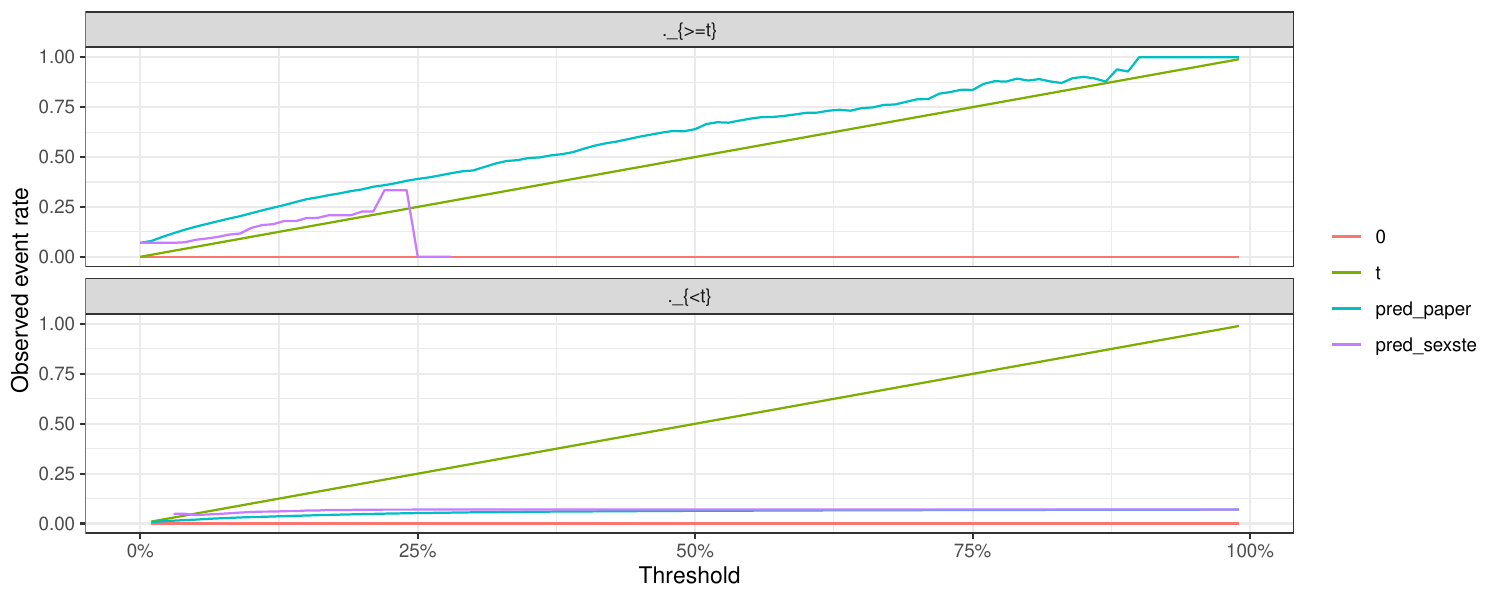}
    \caption{Calibration curve  for GUSTO-I.}
    \label{fig_1_cal}
\end{figure}

\begin{figure}[ht]
    \centering
    \includegraphics[width=1.3\textwidth]{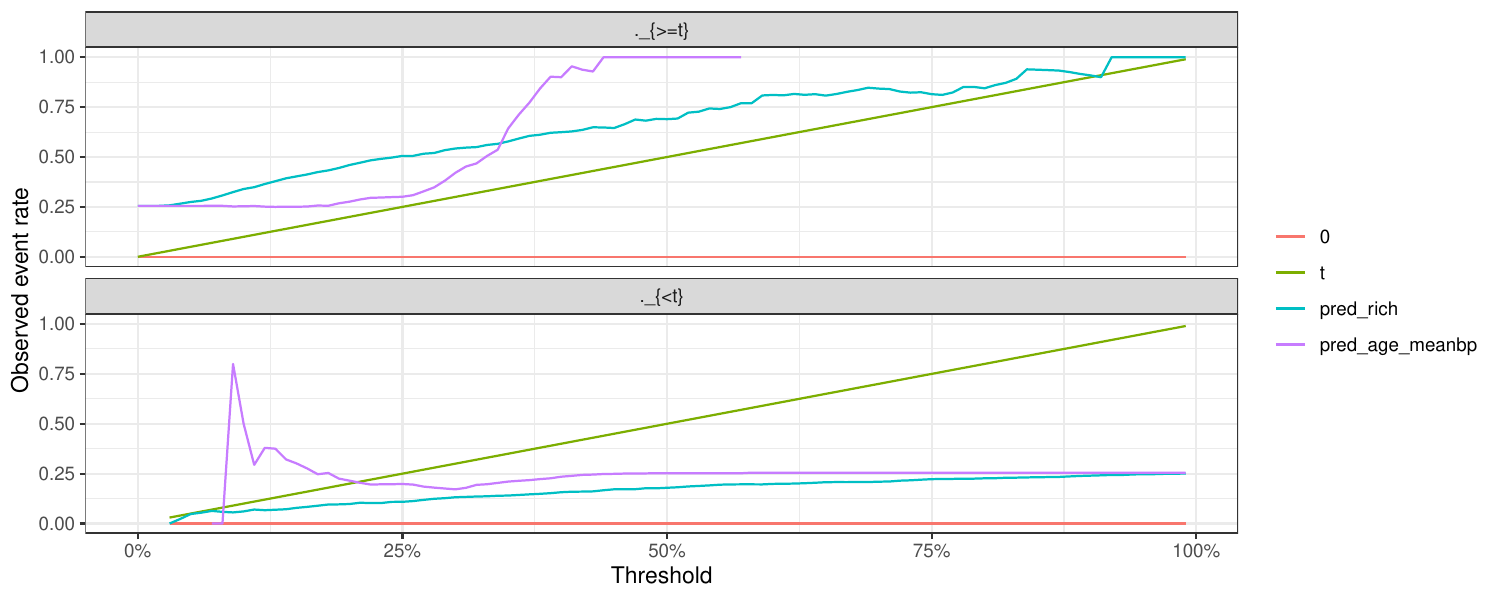}
    \caption{Calibration curve  for SUPPORT.}
    \label{fig_2_cal}
\end{figure}

\section{Mathematical derivations for calibration}\label{A2}
\subsection{Better than treat-none}
Let $s_t>0$, then $\bar Y_{\ge t}:=TP(t)/(TP(t)+FP(t))$ and
\[
\frac{TP(t)}{n}=s_t\,\bar Y_{\ge t},
\qquad
\frac{FP(t)}{n}=s_t\,(1-\bar Y_{\ge t}).
\]
Substituting into $NB(t)=\frac{TP(t)}{n}-\frac{t}{1-t}\frac{FP(t)}{n}$ yields
\begin{align*}
NB(t)
&=s_t\bar Y_{\ge t}-\frac{t}{1-t}s_t(1-\bar Y_{\ge t})
=\frac{s_t}{1-t}\bigl(\bar Y_{\ge t}-t\bigr),
\end{align*}
which proves \eqref{eq_cal_decomp1}.

\subsection{Outperforming treat all}
Let $w:=\tfrac{t}{1-t}$, $t\in(0,1)$. Using \eqref{eq_n_1_n_0} we write
\[
NB(t)-NB_{\mathrm{all}}(t)
=\left(\frac{TP(t)}{n}-w\frac{FP(t)}{n}\right)-\left(\frac{n_1}{n}-w\frac{n_0}{n}\right)
=\frac{-FN(t)+w\,TN(t)}{n},
\]
and
\[
\bar Y_{<t}=\frac{FN(t)}{TN(t)+FN(t)}.
\]
As $FN(t)=n(1-s_t)\bar Y_{<t}$ and $TN(t)=n(1-s_t)(1-\bar Y_{<t})$, we get
\[
NB(t)-NB_{\mathrm{all}}(t)
=(1-s_t)\Bigl[w-(1+w)\bar Y_{<t}\Bigr]
=\frac{1-s_t}{1-t}\,(t-\bar Y_{<t}),
\]
as $w=\tfrac{t}{1-t}$ and $1+w=\tfrac{1}{1-t}$. Therefore, for $s_t<1$,
\[
NB(t)>NB_{\mathrm{all}}(t)\quad\Longleftrightarrow\quad \bar Y_{<t}<t.
\]
\section{Better models for net benefit}\label{A3}
\subsection{In terms of calibration}
For each model $m\in\{1,2\}$, define
\[
s_{m,t}:=\frac{TP_m(t)+FP_m(t)}{n},\qquad
\bar Y_{m,\ge t}:=\frac{TP_m(t)}{TP_m(t)+FP_m(t)},\qquad
\bar Y_{m,<t}:=\frac{FN_m(t)}{TN_m(t)+FN_m(t)}.
\]
Then
\[
\frac{TP_m(t)}{n}=s_{m,t}\bar Y_{m,\ge t},\qquad
\frac{FP_m(t)}{n}=s_{m,t}\bigl(1-\bar Y_{m,\ge t}\bigr),
\]
and hence
\[
NB_m(t)=\frac{TP_m(t)}{n}-\frac{t}{1-t}\frac{FP_m(t)}{n}
=\frac{s_{m,t}}{1-t}\,(\bar Y_{m,\ge t}-t).
\]
Moreover, with prevalence $\pi=n_1/n$,
\[
\pi=s_{m,t}\bar Y_{m,\ge t}+(1-s_{m,t})\bar Y_{m,<t},
\qquad
NB_{\mathrm{all}}(t)=\frac{\pi-t}{1-t},
\]
so
\[
NB_m(t)=NB_{\mathrm{all}}(t)+\frac{1-s_{m,t}}{1-t}\,(t-\bar Y_{m,<t}).
\]

\section{Examples for moderate calibration}
\subsection{Moderate calibration does not guarantee subgroup-specific nonharmfulness}
\label{app:moderate-calibration-subgroups}

The non-harmfulness guarantee of moderate calibration is an overall
statement. It does not imply that the model is non-harmful in every subgroup.

Consider a population consisting of 50 men and 50 women, and suppose the decision
threshold is \(t=0.5\). The model assigns only two possible predicted risks,
\(p=0.8\) and \(p=0.2\). Let the population be as follows:
\[
\begin{array}{c|c|c|c}
\text{subgroup} & p_i & n
& q_i \\ \hline
\text{men}   & 0.8 & 10 & 0.4 \\
\text{women} & 0.8 & 40 & 0.9 \\
\text{men}   & 0.2 & 40 & 0.1 \\
\text{women} & 0.2 & 10 & 0.6
\end{array}
\]
Then the model is moderately calibrated in the overall population, because
\[
\mathbb{E}(\bar{Y}_{i:p_i=0.8})
=
\frac{10\cdot 0.4+40\cdot 0.9}{50}
=0.8,
\]
and
\[
\mathbb{E}(\bar{Y}_{i:p_i=0.2})
=
\frac{40\cdot 0.1+10\cdot 0.6}{50}
=0.2,
\]
using the definition of moderate calibration in equation \eqref{equation_mod_cal}.

At threshold \(t=0.5\), all patients with \(p=0.8\) are classified as
positive and all patients with \(p=0.2\) as negative. Overall,
\[
\mathbb{E}(\bar Y_{\ge 0.5})=\mathbb{E}(\bar{Y}_{i:p_i=0.8})=0.8>0.5,
\]
so the model is better than treat-none. Also,
\[
\mathbb{E}(\bar Y_{< 0.5})=\mathbb{E}(\bar{Y}_{i:p_i=0.2})=0.2<0.5,
\]
so the model is better than treat-all. Hence the model is nonharmful in the
overall population.

However, the model is harmful within both sex subgroups, although in different
ways. Among men classified positive,
\[
\mathbb{E}_{men}(\bar Y_{\ge 0.5})=0.4<0.5,
\]
so the model is worse than treat-none among men. Among women classified
negative,
\[
\mathbb{E}_{women}(\bar Y_{< 0.5})=0.6>0.5,
\]
so the model is worse than treat-all among women.

Thus, moderate calibration can guarantee nonharmfulness on average in the
target population while masking subgroup-specific harm. 
\subsection{Moderate calibration is sufficient but not necessary}
\label{app:moderate-calibration-not-necessary}

Moderate calibration is sufficient for non-harmful decision making in expectation,
but it is not necessary. A model can fail to be moderately calibrated and still
have net benefit at least as high as treat-none and treat-all.

Consider a population of \(100\) patients. The model assigns only two possible
predicted risks, \(p_i=0.8\) and \(p_i=0.2\), and suppose the true risks are as
follows:
\[
\begin{array}{c|c|c}
p_i & n & q_i \\ \hline
0.8 & 50 & 0.9 \\
0.2 & 50 & 0.1
\end{array}
\]
This model is not moderately calibrated, because
\[
\mathbb{E}(\bar{Y}_{i:p_i=0.8})=0.9\neq 0.8,
\]
and
\[
\mathbb{E}(\bar{Y}_{i:p_i=0.2})=0.1\neq 0.2.
\]
Thus, the definition of moderate calibration in equation
\eqref{equation_mod_cal} is not satisfied.

Nevertheless, the model is non-harmful relative to the default strategies. For
thresholds \(0.2<t\le 0.8\), all patients with \(p_i=0.8\) are classified as
positive and all patients with \(p_i=0.2\) as negative. Hence
\[
\mathbb{E}(\bar Y_{\ge t})
=
\mathbb{E}(\bar{Y}_{i:p_i=0.8})
=
0.9
\ge t,
\]
so the model is not worse than treat-none. Similarly,
\[
\mathbb{E}(\bar Y_{<t})
=
\mathbb{E}(\bar{Y}_{i:p_i=0.2})
=
0.1
\le t,
\]
so the model is not worse than treat-all.

For \(t\le 0.2\), all patients are classified as positive, so the model
coincides with treat-all. In this example, the overall event rate is
$\pi=0.5$.
Hence, for \(t\le 0.2\),
\[
NB_{\mathrm{all}}(t)=\frac{\pi-t}{1-t}
=
\frac{0.5-t}{1-t}
\ge 0,
\]
so treat-all, and therefore the model, is not worse than treat-none.

For \(t>0.8\), all patients are classified as negative, so the model
coincides with treat-none. Since \(t>0.8>\pi=0.5\),
\[
NB_{\mathrm{all}}(t)=\frac{0.5-t}{1-t}<0,
\]
so treat-none, and therefore the model, is not worse than treat-all.
Thus, in this example, the model is non-harmful relative to both default
strategies across all thresholds, despite not being moderately calibrated.
\end{document}